\documentclass[aps,prb,floatfix,twocolumn,superscriptaddress,showpacs,epsf,groupedaddress,longbibliography]{revtex4-2}
\usepackage[utf8]{inputenc} 
\usepackage{amsmath,amssymb,amsfonts}
\usepackage{epstopdf}
\usepackage{epsfig} 
\usepackage{graphicx}
\usepackage{subfigure}
\usepackage{import}
\usepackage{color}
\usepackage{url}
\usepackage{overpic}
\usepackage[utf8]{inputenc}
\usepackage{soul}
\usepackage[final]{changes}

\allowdisplaybreaks

\usepackage[makeroom]{cancel}
\usepackage[colorlinks=true,linkcolor=cyan,citecolor=magenta]{hyperref}

\makeatletter
\renewcommand\tableofcontents{\@starttoc{toc}}
\makeatother
\def\bcen{\begin{center}}
\def\ecen{\end{center}}

\def\a{\alpha}       \def\b{\beta}      \def\d{\delta} 
          \def\th{\theta}

\def\t{\tau}           
        \def\o{\omega}   
       \def\D{\Delta}      
                     
        \def\O{\Omega}

\def\aa{{\V \a}}

\def\=={\equiv}

\def\qed{\raise1pt\hbox{\vrule height5pt width5pt depth0pt}}

\def\cG0{{\cal G}_0} 
\def\cG{{\cal G}}

\def\bq{{\bf q}}

 \def\=={\equiv}

\def\Tr{{\rm Tr}}
 \def\ep0{\epsilon_{p}} \def\ed0{\epsilon_{f}}

\def\be{\begin{equation}}
\def\ee{\end{equation}}

\def\ac{a^{\dagger}}
\def\aa{a^{\phantom{\dagger}}}

\newcommand{\ket}[1]{|{#1}\rangle}

\newcommand{\bd}[1]{\mathbf{#1}}
\newcommand{\bs}[1]{\boldsymbol{#1}}
\newcommand{\quave}[1]{\langle{#1}\rangle}

\newcommand{\mean}[1]{\langle #1 \rangle}
\DeclareMathOperator{\tr}{tr}

\def\bx{\mathbf{x}}

\newcommand{\dex}[1]{\left.\D \mathbb{X}_{#1}^2\right._\rho }
\newcommand{\dep}[1]{\left.\D \mathbb{P}_{#1}^2\right._\rho }

\begin{document}

\author{Giacomo Mazza}
\affiliation{Department of Physics ``E. Fermi'' University of Pisa, Largo B. Pontecorvo 3, 56127 Pisa, Italy}

\author{Costantino Budroni}
\affiliation{Department of Physics ``E. Fermi'' University of Pisa, Largo B. Pontecorvo 3, 56127 Pisa, Italy}

\title{Entanglement detection in quantum materials with competing orders }

\begin{abstract}
We investigate entanglement detection in quantum materials through  criteria based 
on the simultaneous suppression of collective matter excitations. 
Unlike other detection schemes, these criteria can be applied to continuous and unbounded variables.
By considering a system of interacting dipoles on a lattice, we show 
the detection of collective entanglement arising from two different physical mechanisms,
namely,  the ferroelectric ordering and the dressing 
of matter degrees of freedom by light. 
In the latter case, the detection shows the formation of a collective entangled 
phase not directly related to spontaneous symmetry breaking. 
These results open a new perspective for the entanglement characterization 
of competing orders in quantum materials, 
and have direct application to quantum paraelectrics with large polariton splittings.
\end{abstract}

\maketitle

{\it Introduction.---}
Entanglement plays a pivotal role in characterizing collective behavior in quantum matter.
Collective entanglement is naturally linked to the presence of quantum 
critical behavior associated with spontaneous symmetry breaking~\cite{Osterloh_Fazio_NATURE2002,Vidal_spin_squeezing_QPT_PRA2004,
Hauke_QFI_susceptibility_NATPHYS2016,OsbornePRA2002, RoscildePRL2004}.
However, entanglement can also become a distinctive feature of systems 
in which quantum collective behavior remains elusive 
due to the absence of explicit symmetry breaking. This is evident in 
cases such as spin liquids, Mott insulators, or other types of topological 
orders~\cite{Jiang_topological_entanglement_entroopy_NP2012,Savary_QSL,
WalshPNAS2021,bellomia2023Mott_entanglement}.
This becomes even more relevant in quantum materials characterized 
by multiple orders emerging from competing collective behaviors~\cite{Giustino_quantum_materials_roadmap2021,basov_quantum_materials_2017}. 

Entanglement detection is now achievable across various quantum systems and degrees of freedom, 
employing a wide array of methods~\cite{Guhne2009,Horodecki2009,Friis2019}. 
These methods range from the exact reconstruction of the density matrix for few-particle systems to entanglement witnesses based solely on measuring a limited number of collective variables. The latter approach is particularly pertinent for many-body systems, especially materials, which often lack the level of tunability found in simpler systems; see Ref.~\cite{Laurell_2024} for a review on entanglement detection in condensed matter systems and Ref.~\cite{Frerot_2023} for a review on detecting quantum correlations 
in many-body systems. In recent years,  quantum Fisher information (QFI) \cite{Helstrom1969, BraunsteinPRL1994} associated with a collective variable has emerged as a powerful technique for detecting entanglement  \cite{PezzePRL2009, HyllusPRA2012, Toth_2014}. 
In particular, QFI can be extracted from the measurement of 
dynamical susceptibilities achievable through state-of-the-art spectroscopic techniques~\cite{Wiesniak_2005, Hauke_QFI_susceptibility_NATPHYS2016,
Mathew_QFI_spin_chain_PRL2020, LaurellPRL21,
Baykusheva_neq_fisher_PRL2023, Hales_light_driven_entanglement_NATCOMM2023,paschen_QFI_strange_metal_2024}.
QFI establishes a connection between entanglement and large collective quantum fluctuations, serving as an entanglement witness when it surpasses a certain threshold, namely its maximum value across all separable states. However, its applicability is confined to collective excitations with bounded spectra, such as spin degrees of freedom. Yet, collective excitations within quantum materials typically encompass a broad range 
of observables with more complex spectra. Consequently, the practical 
implementation of QFI detection schemes can be constrained by the determination 
of the bound as it happens,  e.g, 
in the cases of continuous~\cite{GessnerPRA2016} or non-Hermitian variables~\cite{ren2024witnessing_2024}.
A way around this problem is suggested by the observation that entanglement is also revealed by small quantum fluctuations. Famous examples include spin-squeezing phenomena \cite{Sorensen2001, SorensenPRL2001, TothPRA2009, VitaglianoPRL2011} as well as  the Einstein-Podolsky-Rosen (EPR) argument~\cite{EPR1935} with the associated vanishing uncertainties for the center of mass position and relative momentum \cite{ReidPRA1989,ReidRMP2009}.

Here, we apply criteria based on the 
simultaneous suppression of collective excitations
to the detection of entanglement emerging from 
competing orders in quantum matter.
We consider a quantum paraelectric as a prototype 
system in which collective behavior can emerge 
from the interplay between ferroelectric quantum criticality~\cite{Muller_STO_1979,Rowley_STOcriticality_NatPhy2014,
Rischau_NatPhys2017,Narayan_multiferroic_quantum_criticality2019}, 
and the collective dressing by strong light-matter coupling at equilibrium~\cite{garcia_vidal_ciuti_ebbesen_2021,cavity_quantum_material_2022,
jarc_nature_2021,appugliese_science_2022,giacomo_antoine_sxi,andolina_nogo,passetti_light_matter_entanglement_2023}.
Entanglement is witnessed by the simultaneous suppression of the 
collective excitations which anti-correlate, respectively, with the incipient 
ferroelectric ordering and the formation of polaritons. 
These two different types of detection schemes are intuitively related to two
different physical mechanisms of formation of entanglement.
In the former case, entanglement is a direct consequence 
of intrinsic quantum critical behavior. 
In the latter case, entanglement is transferred, in thermal equilibrium, 
from photons to matter degrees of freedom. We show that the witness 
associated with ferroelectricity is made entanglement-blind by 
the light-matter interaction and vice versa,
thus highlighting the competing nature of the two origins of collective 
entanglement.
The detection scheme is completely general and can be 
applied to any system whose relevant excitations 
are described by pairs of conjugate variables.

\begin{figure}
\includegraphics[width=0.95\columnwidth]{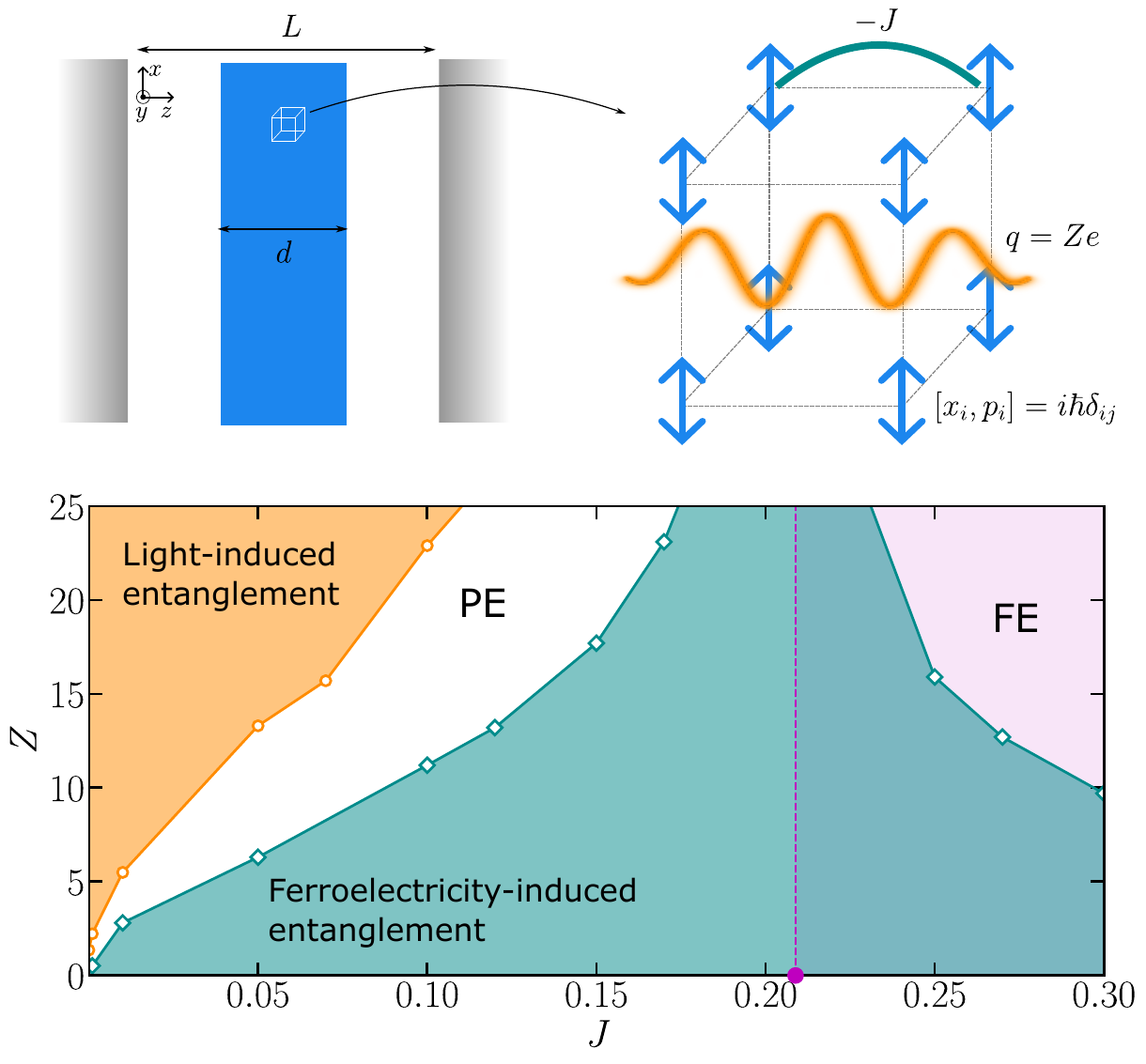}
\caption{Top: (Left) Schematic representation of a slab 
of quantum paraelectric of thickness $d$ in a planar 
cavity of length $L$. (Right) cubic lattice of dipoles described by 
pairs of conjugate variables with nearest neighbor 
ferroelectric coupling $(-J)$ and collective coupling with light in 
the cavity parametrized by the effective charge $q=Ze$, see Eq.~\eqref{eq:Hgeneral}.
Bottom: Entanglement phase diagram in the $Z-J$ plane at $T=0$. 
The colored regions indicates the detection regions for entanglement 
between the dipoles induced, respectively, by the ferroelectric 
coupling (darkcyan) and the collective light-matter coupling (orange). 
The dot and the dashed line indicate the quantum phase transition between the 
paraelectric (PE) and ferroelectric (FE) phases.}
\label{fig:fig0}
\end{figure}

{\it Model and entanglement criteria.---} 
We consider a system of $N$  one-dimensional quantum oscillators of 
mass $m$ localized on the sites of a three-dimensional cubic lattice.  
On each site, the dipoles oscillate along the $x$-direction 
and are described by pairs of conjugate variables $\left[x_i,p_j \right] =  i \hbar  \d_{ij}$. 
The Hamiltonian for independent dipoles reads
\begin{equation}
H_0 = \sum_i H_{0,i} = \sum_{i} \frac{p_i^2}{2 m} + \sum_i V_i(x_i)
\label{eq:bare_hamiltonian}
\end{equation}
where $V_i(x_i) = V(x_i-R_i^x)$ is the on-site potential
centered at the $i$-th lattice site identified by the site vector $\bd{R}_i$.
We consider a quartic form of the potential,
\begin{equation}
V(x) =  m \o_0^2 x^2 \left( \frac{1}{2}+ k^2 \frac{m \o_0}{\hbar} x^2 \right),
\label{eq:quartic_potential}
\end{equation}
where $\o_0 $ is the harmonic frequency and $k$
parametrizes the anharmonic part of the potential. 

We supplement the model with two independent interactions: 
(i) an intrinsic nearest-neighbor dipole-dipole interaction and (ii) the collective light-matter coupling 
between dipoles and the vacuum fluctuations of the electromagnetic fields (see Fig.~\ref{fig:fig0}).
These interactions are parametrized, respectively, by 
a dimensionless ferroelectric coupling, $J>0$, and the effective 
charge of the dipoles, $q=Ze$, with $e$ being the elementary charge.
The full Hamiltonian becomes
\begin{equation}
H = \sum_{i} \frac{\left( p_i + Z e \hat{A}_i \right)^2}{2m} 
+ V_i(x_i) - \frac{J}{2} m \o_0^2 \sum_{\quave{ij}} x_i x_j + H_{em}. 
\label{eq:Hgeneral}
\end{equation}
Here, $\hat{A}_i = \sum_{\mu} A_{\mu,i} \left( \ac_{\mu} + \aa_{\mu} \right)$ is the 
vector potential operator,  with $\ac_{\mu}$ and $\aa_{\mu}$ being photon creation and annihilation 
operators, and  $H_{em} = \sum_{\mu} \hbar \O_{\mu} \ac_{\mu} \aa_{\mu}$ is the free 
photon Hamiltonian. $\O_\mu$ is the frequency of the photon modes, 
with $\mu$ running over all the modes confined between two infinite 
parallel mirrors at distance $L$~\ref{app:light-dressed}. 
The index $i$ runs over all the $N$ dipoles. 
In the following, we assume the thermodynamic limit $N \to \infty$,
with the cubic lattice of finite thickness $d$ and the infinite 
square lattice in the $x$-$y$ plane.

The two interactions act as independent sources of emergent collective behavior.
As a function of $J$, the model describes 
quantum criticality associated with ferroelectric order,~i.e., $\quave{x_i}-R_i^x \neq 0$~\cite{gillisPRB1974,palova_paraelectrics_prb2009,roussev_millis_prb2003}.
At a finite $Z \neq 0$, the dipoles hybridize with light to form collective 
hybrid light-matter excitations dubbed as polaritons.
The interplay between polariton formation and ferroelectricity has recently attracted 
a great deal of attention in relation to the 
possibility of controlling ferroelectricity in the
so-called quantum paraelectrics, such as SrTiO$_3$~\cite{Muller_STO_1979,Rowley_STOcriticality_NatPhy2014}, 
for which the light-matter coupling is particularly strong~\cite{ashida_ferroelectrics_PRX2020,
Shin_quantum_paraelectric_prb2021,Pilar_thermodynamicsof_Quantum2020,latini_sto_pnas2021,
curtis_local_fluctuations_PRR2023}.

We define entanglement with respect to the local partitioning of the Hilbert 
space. A generic state $\rho$ is said to be separable if 
it can be written as a convex combination
of product states:
\begin{equation}
\rho = \sum_\a \lambda_\a \rho_\a^{(1)}\otimes\ldots\otimes \rho_\a^{(N)} 
\label{eq:SEP_def}
\end{equation}
with $\lambda_\a \geq 0$, and $ \sum_\a \lambda_\a=1$.
We denote the associated set as ${\rm SEP}$. 
The extreme points of this set, with respect to convex 
combinations, are pure separable states, 
which take the form of product states, i.e., $\ket{\psi}=\bigotimes_i \ket{\psi_i}$. 
States that are not in ${\rm SEP}$ are said to be entangled.

To detect entanglement, we 
consider a many-body extension of the criterion introduced by 
Duan {\it et al.}~\cite{DuanPRL2000} and Simon~\cite{SimonPRL2000} for two-particle systems 
(see, e.g., Ref.~\cite{vanLoockPRA03}). We specialize it to the case of a periodic lattice, where the collective measurements of position and momentum observables, necessary to detect entanglement, can be represented in terms of reciprocal space and position variables, associated with wave vectors in the first Brillouin zone (BZ). As we discuss below, this representation and the associated physical intuition play a central role in the identification of different origins of the entanglement in different phases of the system.

Starting from the dimensionless local position and momentum variables, 
$X_i := x_i \sqrt{\frac{m \o_0}{\hbar}}$ and 
$P_i := \frac{p_i}{\sqrt{\hbar m \o_0}} $, 
we define reciprocal space position and momentum 
variables as $\mathbb{X}_{\bq} := \frac{1}{\sqrt{N}} \sum_{j} e^{i \bq \bd{R}_j} X_j$ 
and $\mathbb{P}_{\bq} := \frac{1}{\sqrt{N}} \sum_{j} e^{i \bq \bd{R}_j} P_j$, such 
that $\left[ \mathbb{X}_{\bq}, \mathbb{P}_{\bq'} \right] = \d_{\bq,-\bq'}$.
For a generic operator $O$, not necessarily Hermitian, and a state $\rho$, we define the the fluctuation of $O$ on $\rho$ as 
$\Delta O^2_\rho := \mean{O^{\dagger} O}_\rho - \mean{O}_\rho \mean{O^{\dagger}}_\rho $,  
where $\mean{\cdot}_\rho:=\Tr[ \cdot \rho]$ indicates the trace.
If the state is the ground or a thermal state of a given Hamiltonian, 
the fluctuation can be extracted from response functions by the fluctuation-dissipation  theorem~\cite{michele_many_body}
\begin{equation}
\D O^2_{\rho} = \hbar \int_0^{\infty} d \o \left( - \frac{1}{\pi} {\rm Im } \chi (\o) \right) 
\coth \left( \frac{\b \hbar \o}{2} \right)
\label{eq:fluct_diss_theo}
\end{equation}
where $\chi (\o) := \int d t e^{i \o t} \chi(t)$,  with 
$\chi(t) := -\frac{i}{\hbar} \theta(t) \mean{\left[ {\cal O}(t), {\cal O}^{\dagger} \right]}_\rho$ 
and ${\cal O} := O - \mean{O}_{\rho}$, is the response function in the frequency domain,
and $\beta$ is the inverse temperature.

Given position and momentum fluctuations for two wave-vectors
$\bq$ and $\bq'$ the entanglement criterion reads
\begin{equation}
{\rm EW} (\mathbb{X}_{\bq},\mathbb{P}_{\bq'}) :=  \left.\D \mathbb{X}_{\bq}^2\right._\rho + \left.\D \mathbb{P}_{\bq'}^2\right._\rho < 1 
\Rightarrow ~~\rho \notin {\rm SEP},
\label{eq:EPRcriterionq}
\end{equation}
where the symbol ${\rm EW}(\mathbb{X}_{\bq},\mathbb{P}_{\bq'})$ 
indicates the entanglement witness associated with the set of 
$(\mathbb{X}_{\bq},\mathbb{P}_{\bq'})$ operators.
The threshold $1$ is computed by minimizing the fluctuations over all $\rho \in {\rm SEP}$, which, by the concavity of $\Delta O^2_\rho$, is equivalent to minimizing it over all pure product states. There, the total fluctuation is just the sum of local fluctuations for $X_i$ and $P_i$, which obey an uncertainty relation \cite{Busch2014RMP}. 
In order to detect entanglement, we construct witnesses 
using the reciprocal variables 
$\mathbb{X}_{\bq}$ and $\mathbb{P}_{\bq'}$ at different 
wave vectors, i.e., $\bq \neq -\bq'$.
This choice guarantees that   Eq.~\eqref{eq:EPRcriterionq}
indeed defines an entanglement witness, namely, it is 
able to detect at least one entangled state, which is the common eigenstate of $\mathbb{X}_{\bq}$ and
 $\mathbb{P}_{\bq'}$. This state exists because the commutator $\left[\mathbb{X}_{\bq}, \mathbb{P}_{\bq'} \right] = 0$, for $\bq \neq -\bq'$ \footnote{Notice that $\mathbb{X}_{\bq}$ and $\mathbb{P}_{\bq'}$ are normal operators, namely, $[\mathbb{X}_{\bq}, \mathbb{X}_{\bq}^\dagger]=0$, which implies that they can be diagonalized. }. 
Most importantly, 
one can show that, in the absence of symmetry breaking, 
$\left.\D \mathbb{X}_{\bq}^2\right._\rho + \left.\D \mathbb{P}_{\bq}^2\right._\rho
= \left.\D \mathbb{X}_{\bq}^2\right._\rho + \left.\D \mathbb{P}_{-\bq}^2\right._\rho \geq 1$. Therefore,
it would be impossible detect entanglement using $\bq=\pm\bq'$ in the symmetric phase (see \ref{app:ent_crit} for details).
Finally, we stress that the criterion is a sufficient 
condition, and the failure to fulfill Eq.~\eqref{eq:EPRcriterionq} 
does not necessarily mean that the state is separable.

{\it Entanglement at the ferroelectric quantum critical point.---}
We first set $Z=0$ and discuss entanglement detection 
in the ferroelectric model by considering the fully isotropic case $d \to \infty$, i.e., 
infinite thickness  (see Fig.~\ref{fig:fig0}).
We start from the harmonic case, $k=0$ in Eq.~\eqref{eq:quartic_potential}.
In this limit, the Hamiltonian is exactly diagonalized as 
$H = \sum_{\bq} \hbar \o_{\bq} \ac_{\bq} \aa_{\bq}$
with $\hbar \o_{\bq} = \hbar \o_0 \left[{1 - 2 J \sum_{a={x,y,z}} \cos (q_a a)} \right]^{1/2}$
and $\left[ \aa_{\bq'},\ac_{\bq} \right] = \d_{\bq \bq'}$.
$\bq := \left( q_x,q_y,q_z \right) $ is a wave vector within the first  BZ,  
and $a$ is the lattice parameter.
At $J=0$, the spectrum is dispersionless, $\hbar \o_{\bq} = \hbar \o_0$.  
At finite $J$, the frequency of the modes 
close to the BZ boundary, i.e., $\bq=\bs{\pi} := \pi/a(1,1,1)$, increases (mode hardening).
On the contrary, the frequency of the modes close to  the BZ center, i.e., $\bq=\bs{0}$, 
decreases (mode softening). 
For $J \to 1/6$, the $\bq=\bs{0}$ mode completely softens,  
i.e.,  $\o^2_{\bq=0} \to 0$, signaling 
an instability towards a spectrum unbounded from below for $J > 1/6$.

\begin{figure}
\includegraphics[width=\columnwidth]{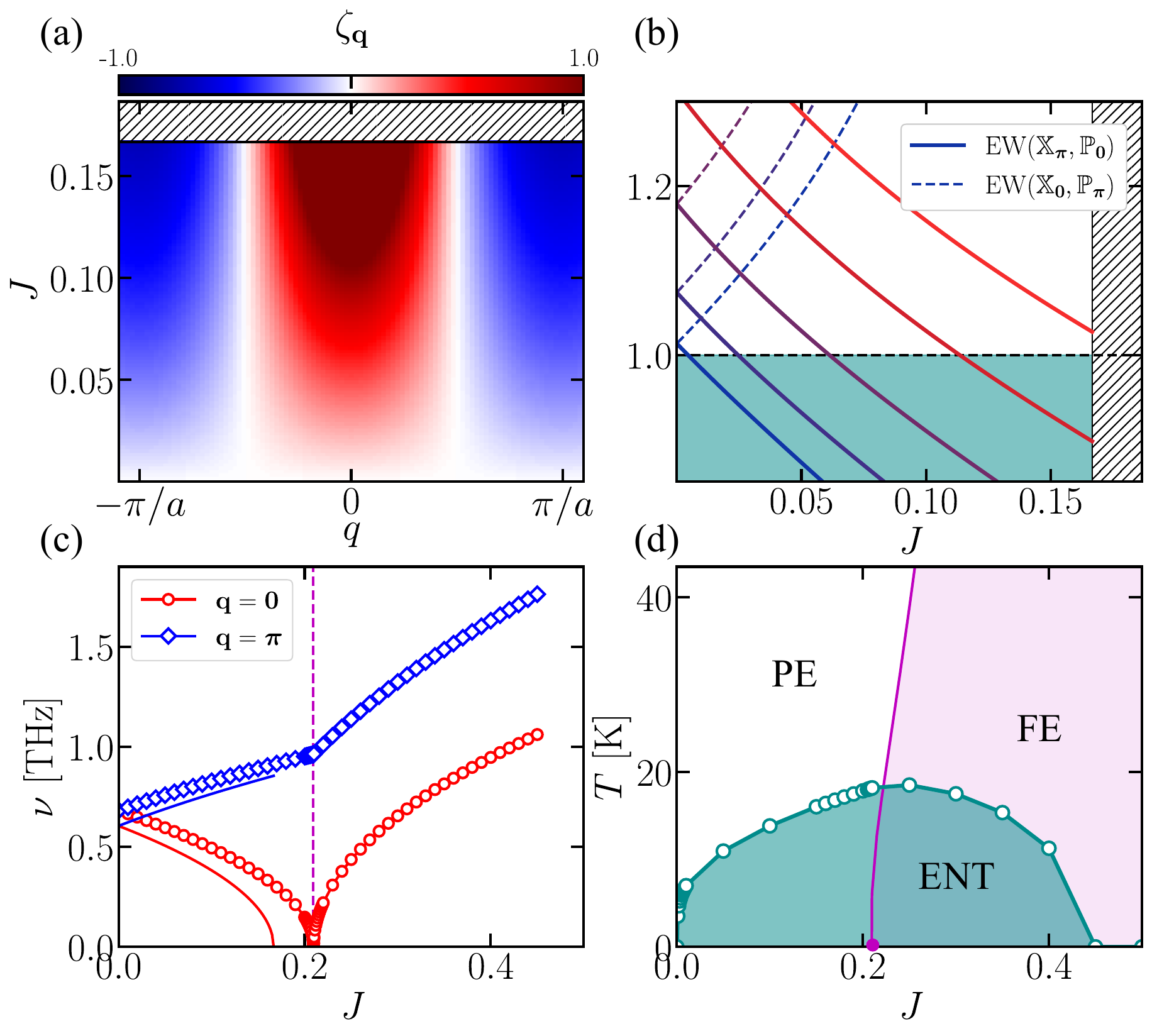}
\caption{Entanglement detection in the purely ferroelectric model.
Top panels: Harmonic potential, $k=0$. 
The hatching in panels (a) and (b) highlights the 
instability of the harmonic model for $J>1/6.$
(a) Squeezing parameter as a function of $\bq=q(1,1,1)$. 
For illustration purposes, $\zeta_{\bq}$ is cut-off between $-1$ and $1$. 
(b) Entanglement witnesses corresponding to the sets 
$(\mathbb{X}_{\bs{\pi}},\mathbb{P}_{\bs{0}})$ (solid lines) and 
$(\mathbb{X}_{\bs{0}},\mathbb{P}_{\bs{\pi}})$  
(dashed lines) for increasing temperature from 
$T_{\rm cold} \simeq 5.8~\rm K$ (blue) to  $T_{\rm hot} \simeq 17.4~\rm K$ (red).
The horizontal dashed line indicates the bound.
Bottom panels:  Anharmonic potential, $k^2=0.05$.
(c) Mode frequencies for $\bq=\bs{0}$ (circles) and $\bq=\bs{\pi}$ 
(diamonds) at $T=0$ across the ferroelectric QCP. 
Solid lines indicates the corresponding values for $k=0$.
(d) Phase diagram in the $J-T$ plane. 
The dot indicates the QCP and the magenta line indicates 
the the FE-PE phase boundary. 
The open circles bounded area indicates the entanglement 
detection region.
}
\label{fig:fig1}
\end{figure}
The mode softening/hardening in different regions of the BZ 
reflects the different energetic costs of parallel/antiparallel 
configuration of dipoles. This observation guides the choice 
of the set $(\mathbb{X}_{\bq},\mathbb{P}_{\bq'})$ 
for entanglement detection
and it is best appreciated 
by looking at the mode-squeezing parameter $ \zeta_{\bq} := 
\log \left( {\D \mathbb{X}_\bq^2}/{\D \mathbb{P}_\bq^2} \right) $,
which measures the strength of the position 
fluctuations with respect to the momentum ones.
A large value of $|\zeta_{\bq}|$ signals the 
tendency to form ordered states with spontaneous 
symmetry breaking signaled by $|\zeta_{\bq}| \to \infty$. 
Specifically, a positive divergence $\zeta_{\bq}$ signals a 
charge instability with $\quave{\mathbb{X}_{\bq}} \neq 0$. 
On the contrary, a negative divergence of $\zeta_\bq$ 
would correspond to the formation of ordered patterns of 
momenta $\quave{\mathbb{P}_{\bq}} \neq 0$ signaling the breaking 
of time reversal symmetry.
In Fig.~\ref{fig:fig1}(a), we plot the mode squeezing parameter 
as a function of $J$ and $\bq$ along the $(1,1,1)$ 
direction of the BZ. 
For $k=0$,  the fluctuation reads
$\D \mathbb{X}_{\bq}^2 = \frac{1}{2 \o_{\bq}} \coth \left( \frac{\b \o_{\bq}}{2} \right)$ 
and $\D \mathbb{P}_{\bq}^2 = \frac{ \o_{\bq} }{2 } \coth \left( \frac{\b \o_{\bq}}{2} \right)$.
At $J=0$, $\D \mathbb{X}_{\bq}^2 = \D \mathbb{P}_{\bq}^2 = 1/2$ 
and $\zeta_{\bq} = 0$ for all $\bq$. 
By increasing $J$,  the softening (hardening) of the frequency $\o_{\bq}$ 
leads to a $\bq-$selective squeezing: Modes at the BZ center become 
momentum squeezed, i.e., $\zeta_{\bq} > 0$, with 
$\zeta_{\bq=0} \to \infty $ for $J \to 1/6$, highlighting the 
ferroelectric instability.
In contrast, modes at the BZ-boundary become position squeezed, i.e., 
$\zeta_{\bq} < 0$. 
Therefore, position and momentum fluctuations  are simultaneously minimized 
[see Eq.~\eqref{eq:EPRcriterionq}] by choosing $\bq$ and $\bq'$, respectively, 
at the boundary and the center of the BZ.
In Fig.~\ref{fig:fig1}(b), we explicitly show the witnesses 
${\rm EW(\mathbb{X}_{\bq=\bs{\pi}},\mathbb{P}_{\bq'=\bs{0}})}$, see Eq.~\ref{eq:EPRcriterionq},  
as a function of $J$ and temperature $T$. 
At $T = 0$, the  entanglement criterion of Eq.~\eqref{eq:EPRcriterionq} is fulfilled for any $J>0$. 
By increasing temperature, the witness is enhanced by thermal fluctuations, 
and the criterion is fulfilled only for a critical coupling $J > J_\star(T)$ 
which monotonically increases with $T$.  Eventually, after a threshold temperature no detection 
is possible in the entire $0 < J < 1/6$ range. 
At the same time, the  complimentary witness 
 ${\rm EW(\mathbb{X}_{\bq=\bs{0}},\mathbb{P}_{\bq'=\bs{\pi}})}$
monotonically increases with $J$ and is never able to detect entanglement.

Turning on a finite $k \neq 0 $, the soft mode instability evolves into a true 
ferroelectric quantum phase transition.
We describe the phase transition by using a Gutzwiller variational ansatz~\cite{gutzwiller_JPB_1992,caleffi_gutzwiller_2020}. 
We find a quantum critical point (QCP) for $J_c \approx 0.209$ 
at the end of a second-order thermal transition line which separates 
the paraelectric (PE) and ferroelectric (FE) phases.
We extract the $\mathbb{X}_{\bq}$ and $\mathbb{P}_{\bq}$ response functions, 
$\chi_{\bq}^{X} (t) := -i \hbar \theta(t) \quave{[\mathbb{X}_{\bq}(t),\mathbb{X}_{-\bq}]}$
and $\chi_{\bq}^{P} (t) := -i \hbar \theta(t) \quave{[\mathbb{P}_{\bq}(t),\mathbb{P}_{-\bq}]} $, 
from the non-equilibrium dynamics in the linear regime~\ref{app:lin_resp}.
Using Eq.~\eqref{eq:fluct_diss_theo}, we compute fluctuations and 
find a dome-like region around the QCP in which the 
set $(\mathbb{X}_{\bq=\bs{0}},\mathbb{P}_{\bq'=\bs{\pi}})$ 
detects entanglement, Fig.~\ref{fig:fig1}(d).
The dome shape of the entanglement detections region is 
understood by observing that, at low temperatures, 
the fluctuations are well approximated by the quasi-harmonic 
expressions, $\D \mathbb{X}_{\bq}^2 \simeq 1/(2 \o_{\bq})$ 
and $\D \mathbb{P}_{\bq}^2 \simeq {\o_{\bq}}/{2} $ 
with mode frequencies defined by averaging over 
the spectral functions $\o_{\bq} := \int_0^\infty d \o A_{\bq}^{X}(\o) \o / \int_0^\infty d \o A_{\bq}^{X}(\o) $, 
with $A_{\bq}^{X}(\o) = -{\rm Im} \chi_{\bq}^{X} (\o) / \pi$.
On the PE side of the transition, the  $\bq=\bs{0}$ and $\bq=\bs{\pi}$
modes closely follow the harmonic results.
By crossing the QCP, the $\bq=\bs{0}$ mode undergoes a 
softening/hardening transition with a cusp-like singularity for $J=J_c,$ 
whereas $\o_{\bs{\pi}}$ monotonically increases with $J$, see Fig.~\ref{fig:fig1}(c).
Therefore, the considerations made for the harmonic case 
on the PE side of the transition get mirrored to the FE side.

\begin{figure}
\includegraphics[width=0.99\columnwidth]{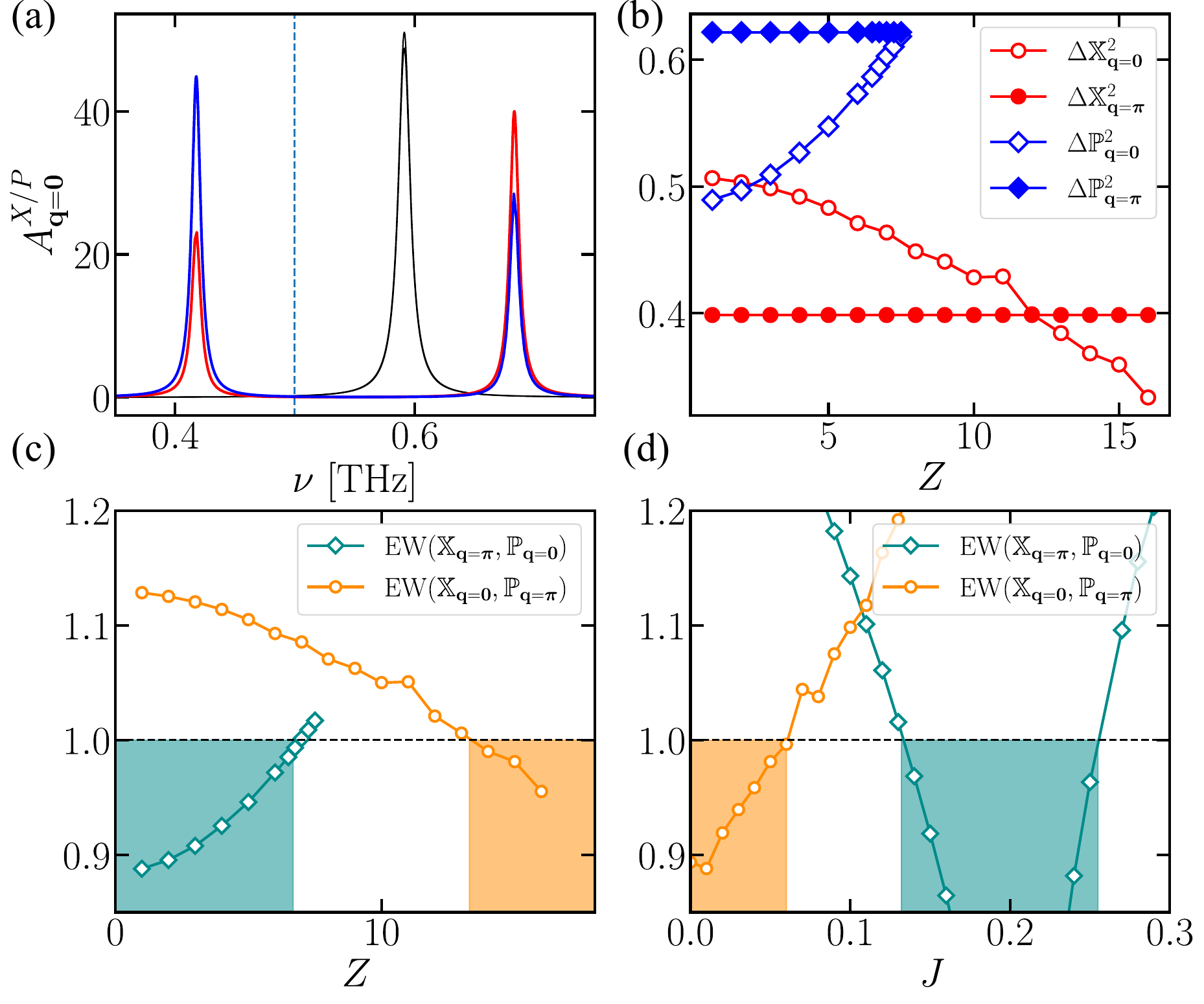}
\caption{ 
Entanglement induced by the light-matter interaction $Z$. 
In all panels $T=0$.
(a) Position (red) and momentum (blue) 
spectral functions for $J=0.05$ and $Z=4.0$, 
compared to the $Z=0$ case (thin black lines).
The dashed line marks the fundamental cavity frequency.
(b) Position (circles) and momentum (diamonds )fluctuations at 
$\bq=\bs{0}$ (open symbols) and $\bq=\bs{\pi}$ 
(solid symbols) as a function of $Z$, and fixed $J=0.05$.
(c) and (d) Entanglement witnesses 
${\rm EW}(\mathbb{X_{\bq=\bs{\pi}}},\mathbb{P}_{\bq=\bs{0}})$
(darkcyan diamonds) and 
${\rm EW}(\mathbb{X_{\bq=\bs{0}}},\mathbb{P}_{\bq=\bs{\pi}})$ 
(orange circles) as a function of $Z$ and fixed $J=0.05$, panel (c), and 
as a function of $J$ and fixed $Z=15$ [panel (d)].
The color code matches the different detection 
regions in the entanglement phase diagram of Fig.~\ref{fig:fig0}.
}
\label{fig:fig2}
\end{figure}

{\it Light-induced entanglement. ---}
We now consider $Z \neq 0$,  and discuss detection 
of entanglement induced by the light-matter coupling.
We compute light-dressed matter response functions 
by including, in the linear response, the dynamics of the self-sourced 
electromagnetic fields~\cite{amelio_prb,flick_acsphotonics_2019}, see also \ref{app:lin_resp}.
To this extent, we set a finite $d=0.2~\mu{\rm m}$ and split the 
site index $i=(n,z)$ 
into in-plane, $n$, and layer, $z=1,\ldots,N_z$, indices. 
We update the definition of the witnesses using
$\mathbb{O}_{\bq=\bs{0}} := \frac{1}{\sqrt{N_{z}}} \sum_{z} O_{\bq_{\parallel}=\bs{0},z} $ 
and $\mathbb{O}_{\bq = \bs{\pi}} =\frac{1}{\sqrt{N_{z}}} \sum_{z} (-1)^{z} O_{\bq_{\parallel}=\bs{\pi},z}$,
for $\mathbb{O}=\mathbb{X},\mathbb{P}$, and $O=X,P$ with $O_{\bq_{\parallel}=\bs{0},z}$ being
the partial Fourier transform in the $x-y$ plane.
Here, we fix the lattice spacing $a=0.5~\rm nm$, leading to $N_z=400$ layers, and 
we set the cavity length to  $L=300~\mu {\rm m}$ in order to have the fundamental 
cavity mode in the THz range. Different choices of parameters do not change 
the qualitative picture discussed below.

In Fig.~\ref{fig:fig2}(a) we report the light-dressed response 
functions for $\mathbb{X}_{\bq=\bs{0}}$ and $\mathbb{P}_{\bq=\bs{0}}$
at $T=0$, compared to the ones for $Z=0$. Panel (b) contains 
the corresponding fluctuations extracted from Eq.~\eqref{eq:fluct_diss_theo}. 
The finite $Z \neq 0$ 
splits the bare resonance into two polariton peaks separated by a gap.
Due to the polariton formation, the homogeneous $\bq=\bs{0}$ fluctuations [Fig.~\ref{fig:fig2}(b)]
get suppressed for the position channel and enhanced for 
the momentum one, ~eventually leading to a change 
of the sign in the squeezing parameter for $\bq=0$.
This behavior can be understood by using a toy model 
of two coupled oscillators with a minimal coupling-like interaction
\ref{app:toy}, and it indicates the tendency of this 
coupling to stimulate breaking of time-reversal 
symmetry~\cite{giacomo_marco_TRSB_2023,mercurio_photon_condensation_van_vleck_2024}.
On the contrary, the staggered fluctuations $ \mathbb{X}_{\bq=\bs{\pi}}$ and $ \mathbb{P}_{\bq=\bs{\pi}}$ 
are not affected at all by the light-matter coupling.
This follows from energetic arguments: 
the frequencies of the dipoles and those of the electromagnetic waves with  
$|\bq| \sim \frac{\pi}{a}$ are orders of magnitude out of resonance. 
Therefore, light dressing is negligible for these modes.  

In Fig.~\ref{fig:fig2}(c),  we report the witnesses for the two 
sets $(\mathbb{X}_{\bq=\bs{\pi}}, \mathbb{P}_{\bq=\bs{0}})$ 
and $(\mathbb{X}_{\bq=\bs{0}}, \mathbb{P}_{\bq=\bs{\pi}})$, 
as a function of $Z$ and fixed $J$. By increasing $Z$, 
the set $(\mathbb{X}_{\bq=\bs{\pi}}, \mathbb{P}_{\bq=\bs{0}})$,  
used to detect entanglement at the ferroelectric QCP, becomes entanglement-blind for $Z>Z_1$. 
The opposite happens for the $(\mathbb{X}_{\bq=\bs{0}}, \mathbb{P}_{\bq=\bs{\pi}})$ 
set which, being entanglement-blind at $Z=0$, starts to detect entanglement for $Z>Z_2$.
Upon inverting the roles of the $(\mathbb{X}_{\bq=\bs{0}}, \mathbb{P}_{\bq=\bs{\pi}})$ and
$(\mathbb{X}_{\bq=\bs{\pi}}, \mathbb{P}_{\bq=\bs{0}})$  sets, the analogous 
behavior is observed by increasing $J$ at fixed $Z$ [see Fig.~\ref{fig:fig2}(d)]. 

{\it Discussion.---}
We summarize the entanglement detection in the phase diagram of Fig.~\ref{fig:fig0}.
Remarkably, the detection region of the $(\mathbb{X}_{\bq=\bs{0}}, \mathbb{P}_{\bq=\bs{\pi}})$ set 
extends down to $J \to 0$, showing that the light-matter coupling acts
as an independent source of entanglement which is not related to the 
ferroelectric one. 
By increasing $J$, the light-induced entanglement 
detection region is pushed to higher values of $Z$,
whereas the presence of the QCP protects the ferroelectric 
entanglement.
From a physical perspective, we can understand these results by noticing that 
photons act as a thermal bath on the dipoles~\cite{fassioli2024controlling_2024},
possibly causing a degradation of the entanglement in the system.
At the same time, due to the strong light-matter coupling, 
entanglement gets transferred from the photon bath 
to the system leading to an entanglement of a different 
origin.
Indeed, it is known that the vacuum of a quantum field
is an entangled state~\cite{Summers1985,Reznik2003} 
and such entanglement can be
transferred, through the mechanism of 
entanglement harvesting \cite{Reznik2003,Reznik2005,Salton2015}, 
to probe two-level detectors as well as more complex physical systems, such as ions in a trap
and cold atoms \cite{RetzkerPRL2005, Ng2008} or quantum dots~\cite{forero2016}.
In this respect, our results show a concrete example of 
entanglement harvesting occurring in the equilibrium state of a quantum material.

The entanglement detection phase diagram 
highlights the starkly different nature of the two types 
of emergent behaviors controlled, respectively, by $J$ and $Z$.
In the first case, entanglement is associated 
with large position fluctuations which eventually 
diverges at the ferroelectric transition. 
On the contrary, light-induced entanglement is 
associated with large momentum fluctuations. 
It is interesting to observe that, even if in our model 
the large momentum fluctuations do not lead to any 
additional symmetry breaking or shift of the ferroelectric QCP, 
the light-matter coupling sensibly modifies the nature 
of the detected entanglement around the QCP, thus 
highlighting the competing nature of the two types of orders.

In summary, we investigated entanglement detection 
in a model of a interacting dipoles in the presence of competing 
orders, linked, respectively, to the ferroelectric QCP and to the collective light-matter coupling. 
We used criteria based on the simultaneous suppression of collective fluctuations 
in position and momentum sampled in different regions of the BZ. 
Collective fluctuations can be extracted from dynamical susceptibilities or 
directly assessed through the measurement of the corresponding 
collective variables~\cite{MengSciAdv2022,EspositoNC2015}.

Our findings directly point to the investigation of entanglement 
in quantum paraelectrics exhibiting significant polariton splitting. 
The detection scheme in combination with the possibility of tuning polaritons~\cite{basov_polaritons_vdw,yixi_polartions_sto2023} 
represents a powerful tool for the control of the collective 
entanglement in these systems.
The construction of different witnesses with different variables
and wave vectors can reveal entanglement of different origin, which 
is not necessarily tied to spontaneous symmetry breaking, potentially 
unlocking the detection in a broad range of quantum materials.

{\it Acknowledgments.---}
The authors thank Giuseppe Vitagliano, Matteo Mitrano 
and Marco Polini for useful discussions. G.M. acknowledges support from the Italian Minister of University and Research (MUR)  under the ``Rita Levi-Montalcini'' program, and from the 
Swiss National Science Foundation through an AMBIZIONE grant (\#PZ00P2\_186145). 
C.B. acknowledges support from the Austrian Science Fund (FWF) through projects 
ZK 3 (Zukunftskolleg) and F 7113 (BeyondC).

\bibliography{references}

\newpage
\onecolumngrid
\appendix

\section{Entanglement criteria}\label{app:ent_crit}
To make our discussion about entanglement self-contained, we provide all the details of the derivation of the entanglement criteria presented in the main text, together with the original references.

Given an operator $A$, its fluctuation on a quantum state $\rho$ is defined as
\begin{equation}
\Delta A^2_\rho := \mean{(A^\dagger-\mean{A^\dagger}_\rho)(A-\mean{A}_\rho)}_\rho = \mean{A^\dagger A}_\rho - \mean{A^\dagger}_\rho \mean{A}_\rho.
\end{equation} 
Notice that, since the operator $A$ is, in general, not Hermitian, we explicitly avoid calling this object a variance.
We also define the symbol $\Delta A_\rho:=\sqrt{\Delta A^2_\rho}$.
Notice that, even if $A$ is not an Hermitian operator, as long as it is normal, i.e., $[A^\dagger,A]=0$,
the interpretation of $\mean{A}$ as an expectation value still holds. In fact, for normal operators the spectral theorem holds: we can still diagonalize it (with complex eigenvalues) and thus make sense of measurements of it. This is the case we consider here.

One can show that the fluctuation is a concave function of the quantum state, namely, for $\rho = \sum_i \lambda_i \rho_i$, with $\{\rho_i\}_i$ quantum states and coefficients $\lambda_i \geq 0$ and $\sum_i \lambda_i= 1$, we have
\begin{equation}\label{eq:concave}
\Delta A^2_{\rho} \geq \sum_i \lambda_i \Delta A^2_{\rho_i}.
\end{equation}
This can be easily shown, with a slight modification of the argument in \cite{HoffmanPRA2003}, as follows
\begin{equation}
\begin{split}
\Delta A^2_{\rho} &= \tr[(A^\dagger-\mean{A^\dagger}_\rho)(A-\mean{A}_\rho)\rho]= \sum_i \lambda_i \tr[(A^\dagger-\mean{A^\dagger}_\rho)(A-\mean{A}_\rho)\rho_i]\\
&=\sum_i \lambda_i \left(\tr[(A^\dagger-\mean{A^\dagger}_\rho)(A-\mean{A}_\rho)\rho_i] +  \mean{A^\dagger}_{\rho_i}\mean{A}_{\rho_i} - \mean{A^\dagger}_{\rho_i}\mean{A}_{\rho_i}\right)\\
&= \sum_i \lambda_i \left(\tr[(A^\dagger A - \mean{A^\dagger}_{\rho_i}\mean{A}_{\rho_i})\rho_i] + \mean{A^\dagger}_{\rho_i}\mean{A}_{\rho_i} - \mean{A^\dagger}_{\rho}\mean{A}_{\rho_i} - \mean{A^\dagger}_{\rho_i}\mean{A}_{\rho} + \mean{A^\dagger}_{\rho}\mean{A}_{\rho}  \right)\\
&=  \sum_i \lambda_i \left( \Delta A^2_{\rho_i}  + |\mean{A}_\rho - \mean{A}_{\rho_i}|^2\right)\\
&\geq \sum_i \lambda_i \Delta A^2_{\rho_i}.
\end{split}
\end{equation}

The concavity property implies that the minimum is achieved on extreme states, i.e., pure. Now consider an operator $A$ on a tensor product $\mathcal{H}=\bigotimes_i \mathcal{H}_i$ defined as a sum of local operators $A=\sum_i \tilde{a}_i$, where $\tilde{a}_i$ is an operator acting on the Hilbert space $\mathcal{H}_i$ and identity everywhere else, e.g., $\tilde{a}_1=a_1\otimes \openone \otimes \ldots \otimes \openone$. Its fluctuation on a product state $\rho=\bigotimes_i \rho_i$ is given by
\begin{equation}\label{eq:prod_dist}
\begin{split}
\Delta A^2_\rho &= \sum_{ij} \tr[\tilde{a}_i^\dagger \tilde{a}_j \bigotimes_k \rho_k] - \sum_{i,j} \tr[a_i^\dagger \rho_i]  \tr[a_j \rho_j]\\
&= \sum_{i\neq j } \left(\tr[a_i^\dagger \rho_i]  \tr[a_j \rho_j] - \tr[a_i^\dagger \rho_i]  \tr[a_j \rho_j]\right) + \sum_i \left(\tr[a_i^\dagger a_i \rho_i]-\tr[a_i^\dagger \rho_i]\tr[a_i \rho_i]\right)\\
&= \sum_i (\Delta a_i^2)_{\rho_i}.
 \end{split}
\end{equation}
Finally, combining Eq.~\eqref{eq:concave}, the fact that all separable states can be written as a convex mixture of pure product states, and Eq.~\eqref{eq:prod_dist}, we have that for any collective variable $A=\sum_i \tilde{a}_i$ defined as above
\begin{equation}\label{eq:min_DA}
\min_{\rho \in {\rm SEP}} \Delta A^2_\rho = \min_{\psi \in {\rm PPROD} }\Delta A^2_\psi = \min_{\{\psi_i\}_i} \sum_i (\Delta a_i^2)_{\psi_i},
\end{equation}
where ${\rm SEP}$ denotes the set of separable states, i.e., states of the form $\rho=\sum_i \lambda_i \sigma_1^{(i)}\otimes \ldots \otimes \sigma_n^{(i)}$, for $\lambda_i\geq 0, \sum_i \lambda_i=1$, ${\rm PPROD}$ the set of pure product states, i.e., $\ket{\psi}=\bigotimes_i \ket{\psi_i}$, $\Delta A^2_\psi$ denotes the fluctuation of $A$ on a global pure state $\ket{\psi}$, and $(\Delta a_i^2)_{\psi_i}$ the fluctuation of the local observable $a_i$ on a local pure state $\ket{\psi_i}$. {\color{black} It is then enough to minimize over all possible collections $\{\psi_i\}_{i=1}^N$ of local states that form the pure product state $\ket{\psi}$.}

We recall the uncertainty relation \cite{Busch2014RMP} {\color{black} for local (i.e., one site) position and momentum operators $X$ and $P$ is}
\begin{equation}
\Delta X \Delta P \geq \frac{1}{2}|\mean{[X,P]}|= \frac{1}{2},
\end{equation}
{\color{black}which combined with $\Delta X, \Delta P \in \mathbb{R}$  and the inequality $(\alpha-\beta)^2=\alpha^2+\beta^2-2\a\b\geq 0$, valid for any $\alpha,\beta\in \mathbb{R}$, gives}
\begin{equation}\label{eq:UR_sum}
{\color{black}\Delta X^2 + \Delta P^2 \geq 2 \Delta X \Delta P \geq |\mean{[X,P]}|=1}.
\end{equation}
{\color{black} Note that the expectation value $\mean{[X,P]}|$ is state independent as $[X,P]=i \openone$.}

{\color{black} Now, consider two collective variables  $A=\sum_i \tilde{a}_i$ and $B=\sum_i \tilde{b}_i$, where  $a_j= e^{i\phi_j} X_j$ and $b_j=e^{i\theta_j} P_j$. We compute}
\begin{equation}\label{eq:min_var_sum}
{\color{black} \min_{\rho \in {\rm SEP}} (\Delta A^2_\rho + \Delta B^2_\rho)=   \min_{\{\psi_j\}_j} \sum_j \left[(\Delta a_j^2)_{\psi_j}+ (\Delta b_j^2)_{\psi_j}\right] = \min_{\{\psi_j\}_j} \sum_j \left[(\Delta X_j^2)_{\psi_j}+ (\Delta P_j^2)_{\psi_j}\right] \geq \min_{\{\psi_j\}_j} \sum_j |\mean{[X_j,P_j]} | = N, }
\end{equation}
{\color{black} where we used Eq.~\eqref{eq:min_DA} for the first equality, the fact that $\Delta a_j^2=\mean{a_j^\dagger a_j}-\mean{a_j^\dagger}\mean{a_j}=\mean{X^2}-\mean{X}^2$ (and similarly for $b_j$) for the second equality, and finally Eq.~\eqref{eq:UR_sum}, which removes the minimization since the result is state-independent.}

{\color{black} Substituting the definition of the reciprocal variable (phases and normalization) one obtains} 
\begin{equation}
\left.\D \mathbb{X}_{\bq}^2\right._\rho + \left.\D \mathbb{P}_{\bq'}^2\right._\rho < 1 
\Rightarrow ~~\rho \notin {\rm SEP}.
\label{eq:EPRcriterion_supp}
\end{equation}
namely, Eq.~\eqref{eq:EPRcriterionq} of the main text. 

{\color{black}
\subsection{Uncertainty relations for equal momentum witnesses}
In this section, we explicitly show that, in the absence of the 
breaking of translation and inversion symmetries, the entanglement 
criterion in Eq.~\eqref{eq:EPRcriterionq} can never be fulfilled by 
pairs of equal momentum reciprocal variables. 
We start by introducing the symmetries. 
Translation symmetry implies that for any pairs of lattice 
sites $i$ and $j$ identified by the lattice vectors $\bd{R}_i$ and $\bd{R}_j$ 
it follows
\begin{equation}
\quave{a_i b_j} = \quave{a_{i+n} b_{j+n}} 
\label{eq:aibj_translation}
\end{equation}
where $a_i$ and $b_j$ represent any local operators
defined on the sites $i$ and $j$, and the indexes $i+n$ and 
$j+n$ label the lattice sites identified, respectively, by 
the lattice vectors $\bd{R}_{i}+\bd{R}_n$ and $\bd{R}_{j}+\bd{R}_n$.
Using the same notation, the inversion symmetry about a given 
lattice site $i=0$ (identified by the null lattice vector $\bd{R}_{i}=0$)
\begin{equation}
\quave{a_0 b_j} = \quave{a_0 b_{-j}},
\label{eq:aibj_inversion}
\end{equation}
where $\bd{R}_{-j}=-\bd{R}_{j} $.
We notice that, Eqs.~\eqref{eq:aibj_translation} and~\eqref{eq:aibj_inversion} are 
always satisfied 
in the paraelectric phase.
On the contrary, Eq.~\eqref{eq:aibj_inversion} does not generally 
hold inside the ferroelectric phase.

We now consider the uncertainty relations for pairs of reciprocal 
variables in the paraelectric phase. 
In this case, since $\mean{\mathbb{X_{\bq}} }=\mean{\mathbb{P_{\bq}}}=0$, $\left.\D \mathbb{X}_{\bq}^2\right._\rho = \quave{\mathbb{X}_{\bq}^{\dagger} \mathbb{X}_{\bq}}_{\rho}$, with $\mathbb{X_{\bq}} = \frac{1}{\sqrt{N}} \sum_i e^{i \bq \bd{R}_i } X_i$, and equivalent expressions for $\left.\D \mathbb{P}_{\bq}^2\right._\rho$.
It is straightforward to see that, due to Eq.~\eqref{eq:aibj_translation} 
$\left.\D \mathbb{X}_{\bq}^2\right._\rho = \left.\D \mathbb{X}_{-\bq}^2\right._\rho$
and $\left.\D \mathbb{P}_{\bq}^2\right._\rho = \left.\D \mathbb{P}_{-\bq}^2\right._\rho$.
To derive the uncertainty relation we write
\begin{equation}
\quave{\left(\mathbb{X}_{\bq}^{\dagger}-i\mathbb{P}^{\dagger}_{\bq'} \right) 
\left(\mathbb{X}_{\bq}+i\mathbb{P}_{\bq'} \right)}_\rho  \geq 0 
\end{equation}
and 
\begin{equation}
\quave{\left(\mathbb{X}_{\bq}^{\dagger}+i\mathbb{P}^{\dagger}_{\bq'} \right) 
\left(\mathbb{X}_{\bq}-i\mathbb{P}_{\bq'} \right)}_\rho  \geq 0,
\end{equation}
from which we obtain
\begin{equation}
\dex{\bq} + \dep{\bq'} \geq |\quave{\mathbb{X}^{\dagger}_{\bq} \mathbb{P}_{\bq'} - \mathbb{P}_{\bq'}^{\dagger} \mathbb{X}_{\bq}}_{\rho} | .
\label{eq:dxq_dpq}
\end{equation}
We notice that Eq.~\eqref{eq:dxq_dpq} precisely follows from 
the uncertainty principle for two general variables $A$ and 
$B$ when $\quave{A}=0$ or $\quave{B} = 0$
\begin{equation}
\D A \D B \geq \sqrt{\left(\frac{\mean{A^\dagger B+B^{\dagger} A}}{2} - {\rm Re}(\mean{A^\dagger}\mean{B})\right)^2 +\left(\frac{\mean{A^\dagger B - B A^{\dagger}}}{2i} - {\rm Im}(\mean{A^\dagger}\mean{B})\right)^2 }.  
\end{equation}
We now expand the right-hand term of Eq.~\eqref{eq:dxq_dpq}
\begin{equation}
\begin{split}
\quave{\mathbb{X}^{\dagger}_{\bq} \mathbb{P}_{\bq'} - \mathbb{P}_{\bq'}^{\dagger} \mathbb{X}_{\bq}}_{\rho} & = \frac{1}{N} \sum_{ij} e^{-i \bq \bd{R}_{i}} e^{i \bq' \bd{R}_j}
\quave{X_i P_j} +  c.c. \\
& =  \frac{1}{N} \sum_{i j} e^{- i (\bq-\bq') \bd{R}_j } e^{-i \bq (\bd{R}_i - \bd{R}_j) } \quave{X_{i-j} P_0} + c.c. \\
& =  \d_{\bq \bq'}  \sum_{n} \left[  e^{- i \bq \bd{R}_n} \quave{X_n P_0} 
- e^{i \bq \bd{R}_n} \quave{P_0 X_n}  \right]
\end{split}
\end{equation}
By inversion symmetry, see Eq.~\eqref{eq:aibj_inversion}, we get
\begin{equation}
\quave{\mathbb{X}^{\dagger}_{\bq} \mathbb{P}_{\bq'} - \mathbb{P}_{\bq'}^{\dagger} \mathbb{X}_{\bq}}_{\rho} =  \d_{\bq \bq'} \sum_{n} e^{- i \bq \bd{R}_n} \quave{\left[ X_n,P_0\right]} = i \d_{\bq \bq'}.
\end{equation}
Finally, the uncertainty relation for equal momentum witnesses reads
\begin{equation}
\dex{\bq} + \dep{\bq} = \dex{\bq} + \dep{-\bq} \geq 1.
\label{eq:sameq_uncertainty}
\end{equation}
The uncertainty relation in Eq.~\eqref{eq:sameq_uncertainty}
can be readily verified by computing fluctuations in the 
purely harmonic model (see main text), 
$\D \mathbb{X}_{\bq}^2 = \frac{1}{2 \omega_{\bq}} \coth \left( \frac{\b \omega_{\bq}}{2} \right)$
and
$\D \mathbb{P}_{\bq}^2 = \frac{\omega_{\bq}}{2 } \coth \left( \frac{\b \omega_{\bq}}{2} \right)$ with $\o_{\bq} = \o_{-\bq}$.
It follows that $ \D \mathbb{X}_{\bq}^2  + \D \mathbb{P}_{\bq}^2 = \D \mathbb{X}_{\bq}^2  + \D \mathbb{P}_{-\bq}^2 = \frac{1}{2} \left( \frac{1}{\o_{\bq}} + \o_{\bq} \right) \coth \left( \frac{\b \omega_{\bq}}{2} \right) \geq 1.$
}

\section{Model of interacting dipoles}\label{app:dynamics}

\subsection{Linear response dynamics}\label{app:lin_resp}
In this section we detail the calculation of the response 
functions using the time-dependent Gutzwiller ansatz.
Our goal is to computed the response functions at wave-vector $\bq$, defined as 
\begin{equation}
\chi_{\bq}^O(t) = -i \theta(t) \quave{\left[ \mathbb{O}_{\bq},\mathbb{O}_{\bq}^{\dagger} \right]} = 
-i \theta(t) \quave{\left[ \mathbb{O}_{\bq},\mathbb{O}_{-\bq} \right]},
\label{eq:chiq}
\end{equation}
for $\mathbb{O} = \mathbb{X},\mathbb{P}$, where 
$\mathbb{O}_{\bq}^{\dagger} = \mathbb{O}_{-\bq}$.
By linear response theory, the response functions can be 
extracted from the unitary dynamics with the Hamiltonian 
supplemented by a small perturbation time-dependent 
perturbation field $\l(t)$.
Specifically, by defining the Hermitian operators 
\begin{align}
\mathbb{O}_{\bq+} := \mathbb{O}_{\bq} +  \mathbb{O}_{\bq}^{\dagger}, 
\qquad  
{\rm and}
\qquad
\mathbb{O}_{\bq-} := -i \left( \mathbb{O}_{\bq} -  \mathbb{O}_{\bq}^{\dagger}  \right),
\end{align}
we introduce the $\bq$- and time-dependent Hamiltonians 
\begin{align}
H_{\bq \pm}(t) &= H + \l(t) \mathbb{O}_{\bq\pm},
\label{eq:Hlinear_response}
\end{align}
and the corresponding time-evolved states  
\begin{align}
\rho_{\bq,\pm}(t) := e^{i \int_0^t d t' H_{\bq \pm}(t')} \rho_0  e^{-i \int_0^t d t' H_{\bq \pm}(t')}.
\end{align}
The time-evolved states $\rho_{\bq,\pm}(t)$ allow the definition of four expectation 
values with functional dependence on the perturbation $\l(t)$,
\begin{align}
O_{\bq}^{\pm \pm} (t) = O_{\bq}^{\pm \pm} \left[ \l(t) \right]
= \Tr \left[ \rho_{\bq,\pm}(t) \mathbb{O}_{\bq \pm} \right].
\end{align}
At linear order in the perturbation $\l(t)$, such a functional dependence 
is encoded in the response function as
\begin{align}
O_{\bq}^{\pm \pm} (t) = \int_{- \infty}^{+\infty} d t' \chi_{\bq}^{\pm \pm} (t-t') \l(t'),
\end{align}
with 
\begin{align}
\chi_{\bq}^{\pm \pm} (t-t') = -i \th(t-t') \quave{\left[ \mathbb{O}_{\bq\pm}(t),\mathbb{O}_{\bq,\pm}(t') \right]}.
\label{eq:chi_pmpm}
\end{align}
From the knowledge of the time-dependent expectation values 
$O_{\bq}^{\pm \pm} (t)$ we determine the response functions, Eq.~\eqref{eq:chi_pmpm}, by Fourier transform
\begin{align}
O_{\bq}^{\pm \pm} (\o) = \chi_{\bq}^{\pm \pm} (\o) \l(\o).
\end{align}
Eventually, the response function, Eq.~\eqref{eq:chiq}, is obtained as
\begin{align}
\chi^O_{\bq}(\o) = \frac{1}{4} \left[ \chi_{\bq}^{++}(\o) + \chi_{\bq}^{--}(\o) 
- i (\chi_{\bq}^{+-}(\o) - \chi_{\bq}^{-+}(\o)) \right].
\end{align}

\subsection{Gutzwiller dynamics}\label{app:gutzwiller}
To study the dynamics of the interacting model of dipoles we use a 
Gutzwiller single-site ansatz
\begin{align*}
\rho(t) = \bigotimes_i \rho_i(t)
\end{align*} 
where $\rho_i(t)$ is a state defined on the Hilbert space of the dipole at site $i$, 
which evolves with an effective single-site Hamiltonian
\begin{align}
H_i(t) = H_{0,i}(t) - x_i J_{eff,i}(t)
\end{align}
with $J_{eff,i}(t) = J m \o_0^2  \sum_{\quave{j}} \Tr \left( \rho_j(t) x_j \right)$ 
where the sum over $j$ is restricted to the nearest neighbor sites of $i$.
In the static limit, this procedure corresponds to the static mean-field 
ansatz which describes the spontaneous symmetry breaking 
at mean-field level. In the time-dependent case, the method is able to capture 
quantum fluctuations on top of the static mean-field. 
In particular, the contributions from all the other sites are encoded 
in the effective field $J_{eff,i}(t)$.
It can be shown that the dynamics is exact in the two limits $\frac{k}{J} \to 0$ 
(harmonic limit) and $\frac{k}{J} \to \infty$ (atomic limit), where $k$ and $J$ 
are, respectively, the anharmonicity parameter and 
the nearest-neighbor coupling constant.

\begin{figure}
\includegraphics[width=0.8\columnwidth]{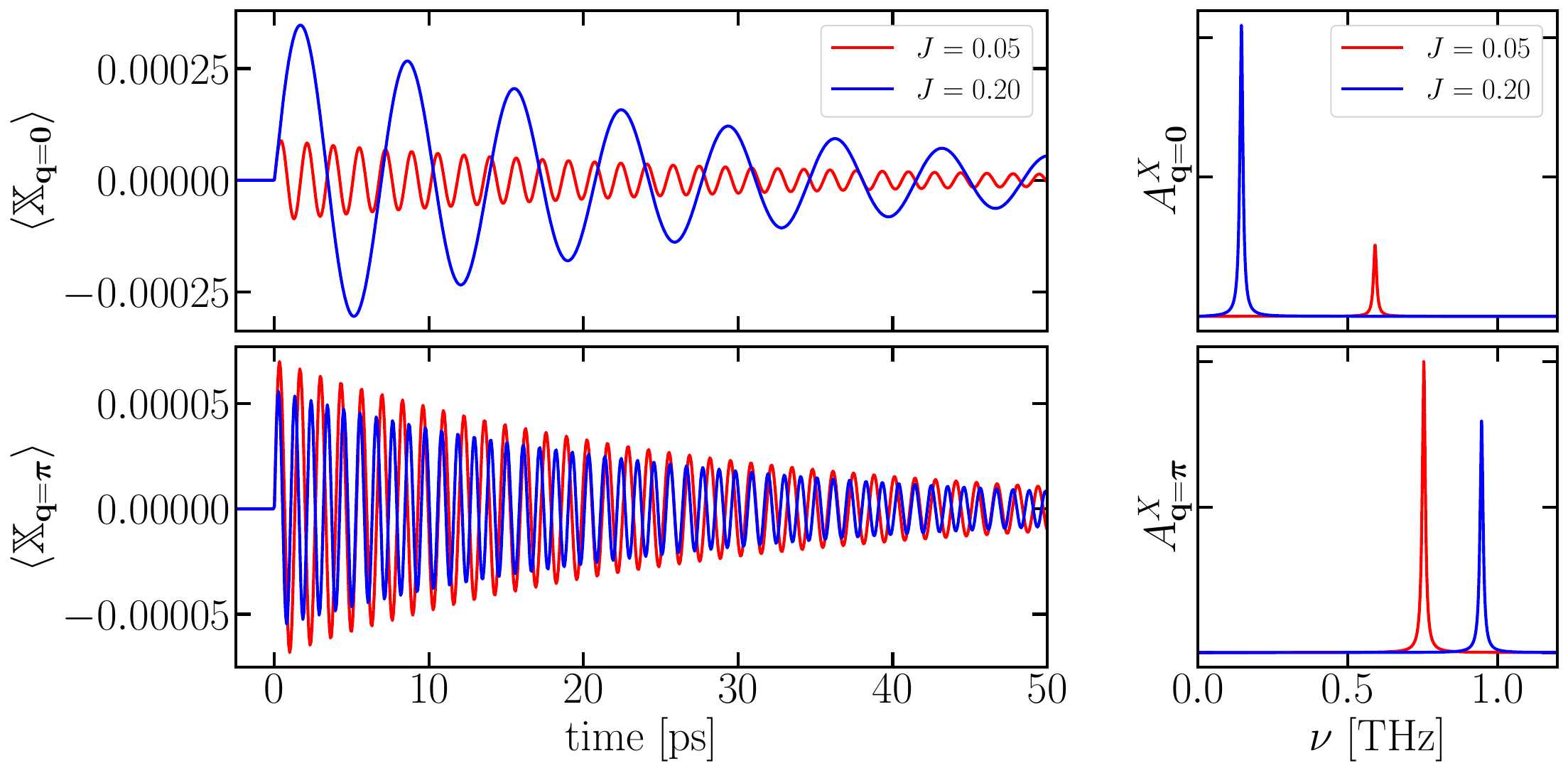}
\caption{Left panels: Linear response dynamics of the position operator for $\bq=\bs=0$ (top)
and $\bq=\bs{\pi}$ (bottom) and increasing value of $J$. 
Right panels: Position response functions obtained by Fourier transform 
of the time signals.}
\label{fig:supp_fig1}
\end{figure}

In order to study the dynamics, we represent the states $\rho_i(t)$
a local local basis sets containing $N_i=10$ eigenstates 
and checked convergence with respect to $N_i$.   
We used a Gaussian perturbation $\l(t) = \l_{0} e^{-\frac{t^2}{\t^2}}$ 
with $\l_0 = 10^{-3} \hbar \o_0$ and $\t=10^{-5}~{\rm ps}$.
In Fig.~\ref{fig:supp_fig1} we show examples of linear response 
dynamics for the position operator with $\bq=\bs{0}$ and the $\bq=\bs{\pi}$ 
wavevectors. 
We extract the response functions by 
evaluating the Fourier transform 
over a time window of  $500~{\rm ps}$. 
In the dynamics, we included a small damping which 
ensures convergence of the Fourier integrals
in the considered time window.

\begin{figure}[h]
\centering
\includegraphics[scale=0.2]{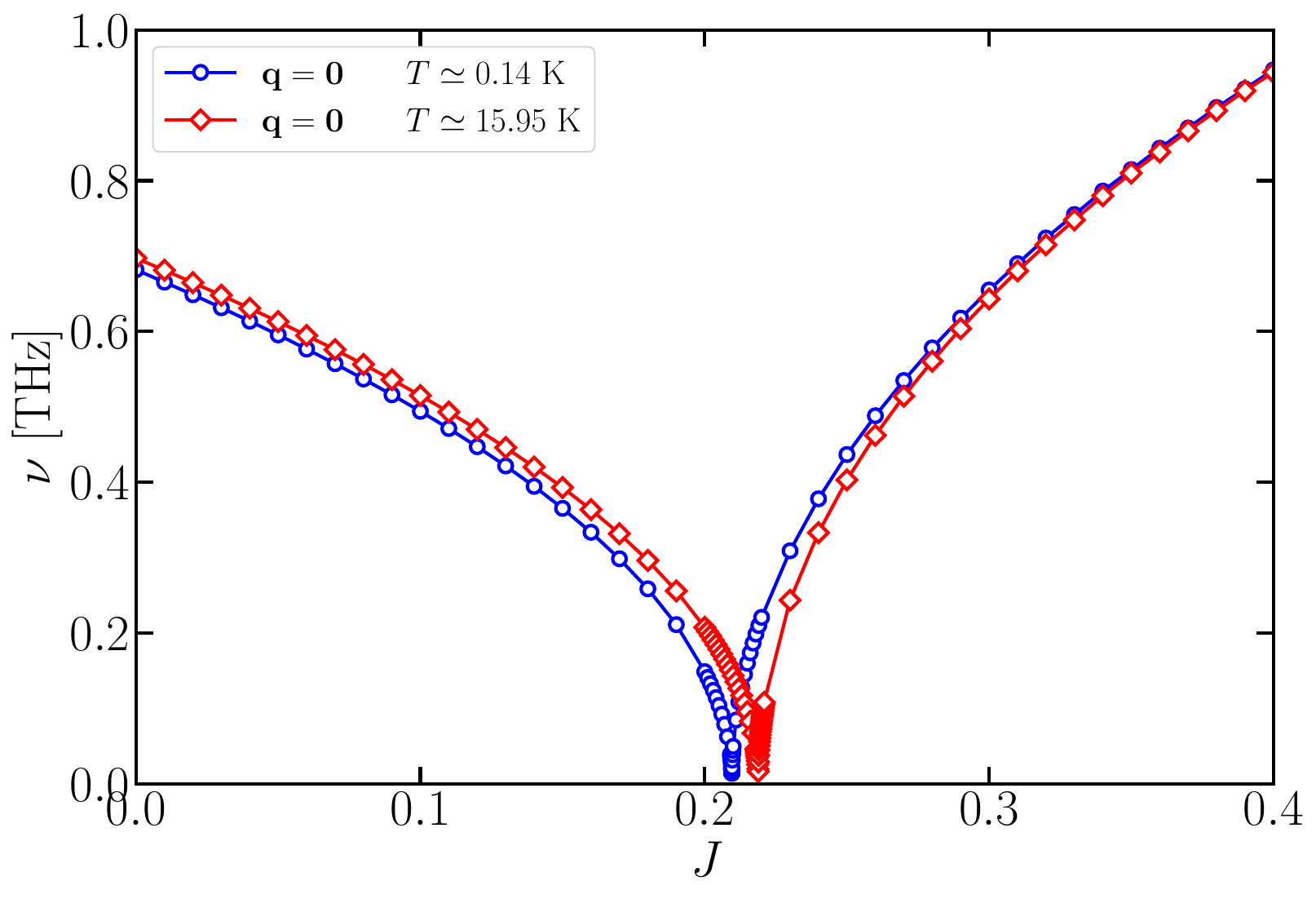}
\caption{Softening of the ferroelectric 
mode at different temperatures and as a function of $J$.}
\label{fig:supp_fig1b}
\end{figure}

In Fig.~\ref{fig:supp_fig1b} we show the frequency of the 
ferroelectric mode extracted from the linear response dynamics 
for different temperatures as a function of $J$. 
The mode softening determines the phase boundary between 
the paralectric and ferroelectric phases.

\newpage
\subsection{Light-dressed response functions}\label{app:light-dressed}
In this section we show details of the calculation of the light-dressed response functions.
We first write the full Hamiltonian of the dipoles interacting with the photon degrees of freedom.
\begin{equation}
H = \sum_{\mu} \hbar \O_{\mu} \ac_{\mu} \aa_{\mu} + 
\sum_{i} \frac{1}{2 m} \left( p_i + Ze \hat{A}_x(x_i) \right)^2 + V_i(x_i) - J m \o_0^2 x^2 \left( \frac{1}{2} + k^2 \frac{m \o_0}{\hbar} x^2 \right)
\label{eq:H_light_matter_full}
\end{equation}
Here, $\hat{A}_x(x_i)$ is the $x-$component of the vector potential 
operator computed at the position of the point-like dipole. 
Specifically, starting from the full vector potential operator defined 
in all the points of the three-dimensional space,
\begin{align}
\hat{\bd{A}}(\bx) = \hat{\bd{A}}(x,y,z) = \bs{x}  \hat{A}_x(x,y,z) + \bs{y} \hat{A}_y(x,y,z) +  \bs{z} \hat{A}_z(x,y,z),
\end{align}
the operator $\hat{A}_x(x_i)$ is defined as
\begin{equation}
\hat{A}_x(x_i) := \int dx dy dz A_x(x,y,z) \d(x-x_i) \d(y-R_{i}^y) \d(z-R_{i}^z).
\end{equation}
Notice that the $y-$ and $z-$ components of the vector potential
do not enter the Hamiltonian as the dipoles oscillates only along 
the $x-$direction. The full vector potential quantized in the volume of the 
cavity reads
\begin{equation}
\hat{\bd{A}}(\bx) = \sum_{\mu} A_{\mu} \left( \bd{u}_\mu (\bx) a_\mu + \bd{u}^*_{\mu}(\bx) \ac_\mu \right)~
\qquad A_\mu =\sqrt{\frac{\hbar^2}{2 \epsilon_0 \O_\mu V }} 
\qquad
\O_\mu = \hbar c |\bd{q}_\mu|
\end{equation}
where the mode functions form a complete set 
of functions which satisfy the wave equation and 
the divergence-less condition
\begin{equation}
\vec{\nabla}^2 \bd{u}_{\mu} + \bq^2_{\mu} \bd{u}_{\mu} = 0
\qquad
\nabla \cdot \bd{u}_{\mu} = 0
\qquad 
\frac{1}{V}\int d\bx ~ \bd{u}_{\mu }^* \cdot \bd{u}_{\mu'} = \d_{\mu  \mu'} 
\label{eq:wave_equation},
\end{equation}
with boundary conditions set by perfectly reflecting mirrors.

\begin{figure}
\includegraphics[width=0.8\columnwidth]{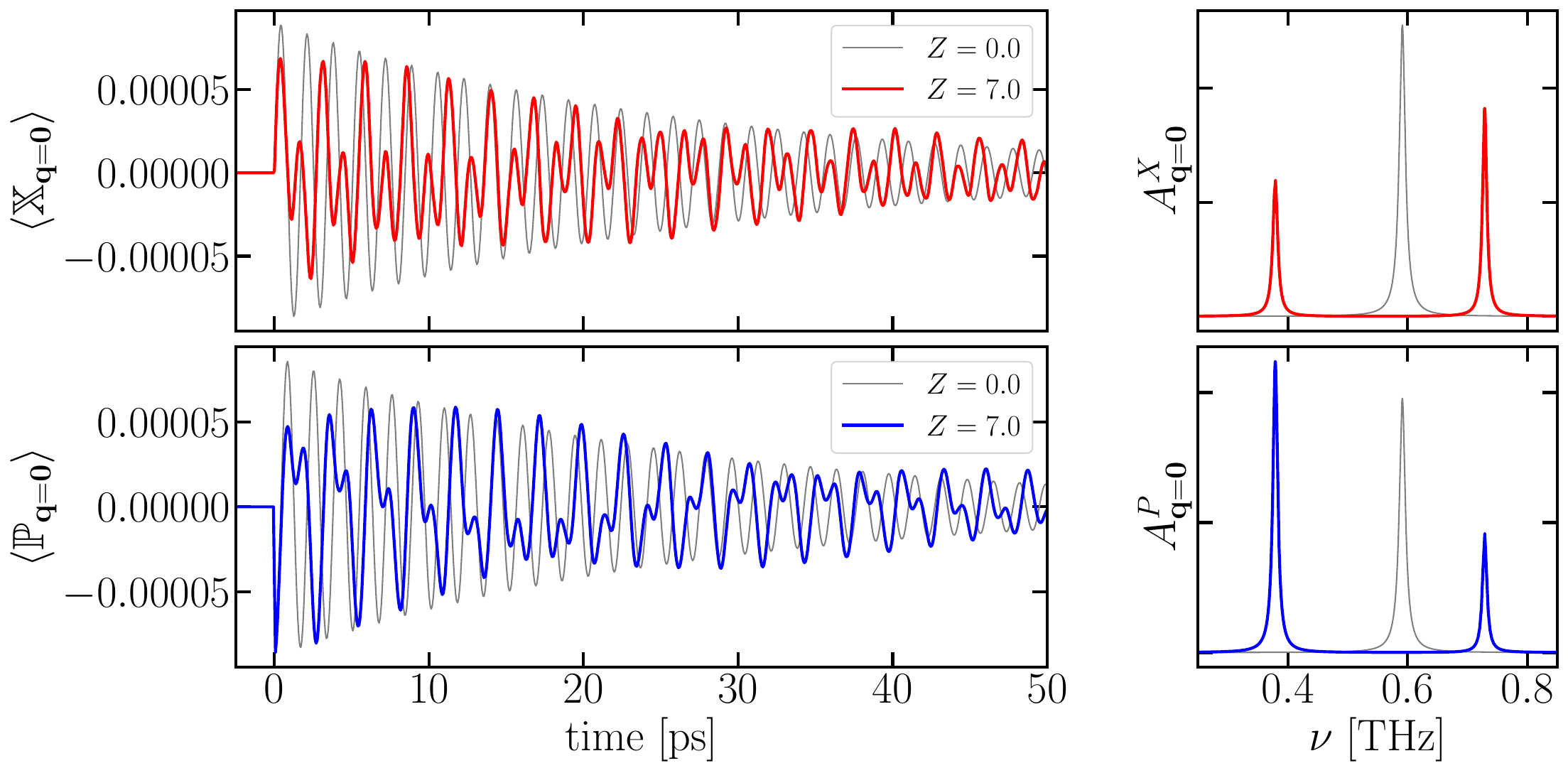}
\caption{Left panels: Linear response dynamics for the homogeneous $\bq=\bs{0}$ 
position (top) and momentum (bottom) for increasing values of the effective charge.
Right panels: Light-dressed position and momentum response function obtained 
by Fourier transform of the time signals.}
\label{fig:supp_fig2}
\end{figure}

We supplement the Hamiltonian with the linear response perturbation, 
as in Eq.~\eqref{eq:Hlinear_response}
\begin{equation}
H_{\bq \pm}(t) = \sum_{\mu} \hbar \O_{\mu} \ac_{\mu} \aa_{\mu} + 
\sum_{i} \frac{1}{2 m} \left( p_i + Ze \hat{A}_x(x_i) \right)^2 + V_i(x_i) - J m \o_0^2 x^2 \left( \frac{1}{2} + k^2 \frac{m \o_0}{\hbar} x^2 \right) + \l(t) \mathbb{O}_{\bq \pm}.
\label{eq:Hdressed_linear}
\end{equation}
We therefore follow Ref.~\cite{amelio_prb} and write the dynamics 
as the coupled dynamics of dipoles in the presence of fields whose evolution 
is governed by the Maxwell equations with the dipoles acting as sources of currents.
Specifically, density matrix of the dipoles evolves with the Hamiltonian
\begin{align}
H_{\bq \pm}\left[ A,t \right] =  
\sum_{i} \frac{1}{2 m} \left( p_i + Ze A_x(x_i) \right)^2 + V_i(x_i) - J m \o_0^2 x^2 \left( \frac{1}{2} + k^2 \frac{m \o_0}{\hbar} x^2 \right) + \l(t) \mathbb{O}_{\bq \pm}.
\label{eq:Hlight_matter_semiclassical}
\end{align}
\begin{align}
i \hbar \partial_t \rho_{\bq \pm} = \left[ H_{\bq \pm}(t), \rho_{\bq \pm} \right].
\label{eq:drhoqdt}
\end{align}
where the field entering  Eq.~\eqref{eq:Hlight_matter_semiclassical} satisfy 
\begin{align}
- {\nabla}^{2} A_x(\bx) 
- \frac{1}{c^2} \frac{\partial^2 A_x(\bx)}{\partial t^2} = \mu_0 {J}_x(\bx).
\label{eq:maxwell_field}
\end{align}
In Eq.~\eqref{eq:maxwell_field}, the current density reads 
\begin{align*}
J_x(\bx) = \sum_i \Tr \left(\rho_{\bq \pm}(t) \hat{J}_{x,i}(\bx) \right),
\end{align*}
being $\hat{J}_{x,i}(\bx)$ the current density operator associated 
with the $i$-th dipole
\begin{align}
\hat{J}_{x,i}(\bx)= \hat{J}_{x,i}(x,y,z) =\frac{Ze}{m} \left( p_i + Z e \hat{A}_{x}(x_i) \right) 
\d(x-x_i) \d(y-R_i^y) \d(z-R_i^z),
\label{eq:point-like-current}
\end{align}
Eventually, the computation of the light-dressed response functions reduces to 
the coupled dynamics of dipoles in the presence of self-sourced fields, 
Eqs.~\eqref{eq:drhoqdt}-\eqref{eq:maxwell_field}. 
The dynamics of the dipoles is solved using the same method described above.
We solve the wave-equation of the field, Eq.~\eqref{eq:maxwell_field}, 
by expanding the field on the quantized mode in the cavity, Eq.~\eqref{eq:wave_equation}. 
In practice, we average the point-like current 
density over a volume $a^3$ around each lattice site and 
assume the vector potential constant within each volume $a^3$.
\begin{align}
J_x(\bx) = \sum_i \frac{1}{a^3} \Tr \left( \rho(t) \hat{J}^x_i \right) 
\theta\left(|x-R^x_i|- \frac{a}{2}\right)
\theta\left(|y-R^y_i|- \frac{a}{2}\right)
\theta\left(|z-R^z_i|- \frac{a}{2}\right)
\end{align}
with 
\begin{align}
\hat{J}^x_i =\frac{Ze}{m} \left[ p_i+Ze A_x(R_i^x,R_i^y,R_i^z) \right]
\end{align}
In all the calculations, we considered an high energy cutoff of $ 0.5~{\rm eV}$ 
on the photon modes and checked convergence with the cutoff.
In Fig.~\ref{fig:supp_fig2}, we show an example of light-dressed linear 
response dynamics for the homogeneous position and momentum perturbations.
Frequency integration of the response function yields the 
position and momentum fluctuations via the fluctuation/dissipation 
theorem.

\begin{figure}
\includegraphics[scale=0.3]{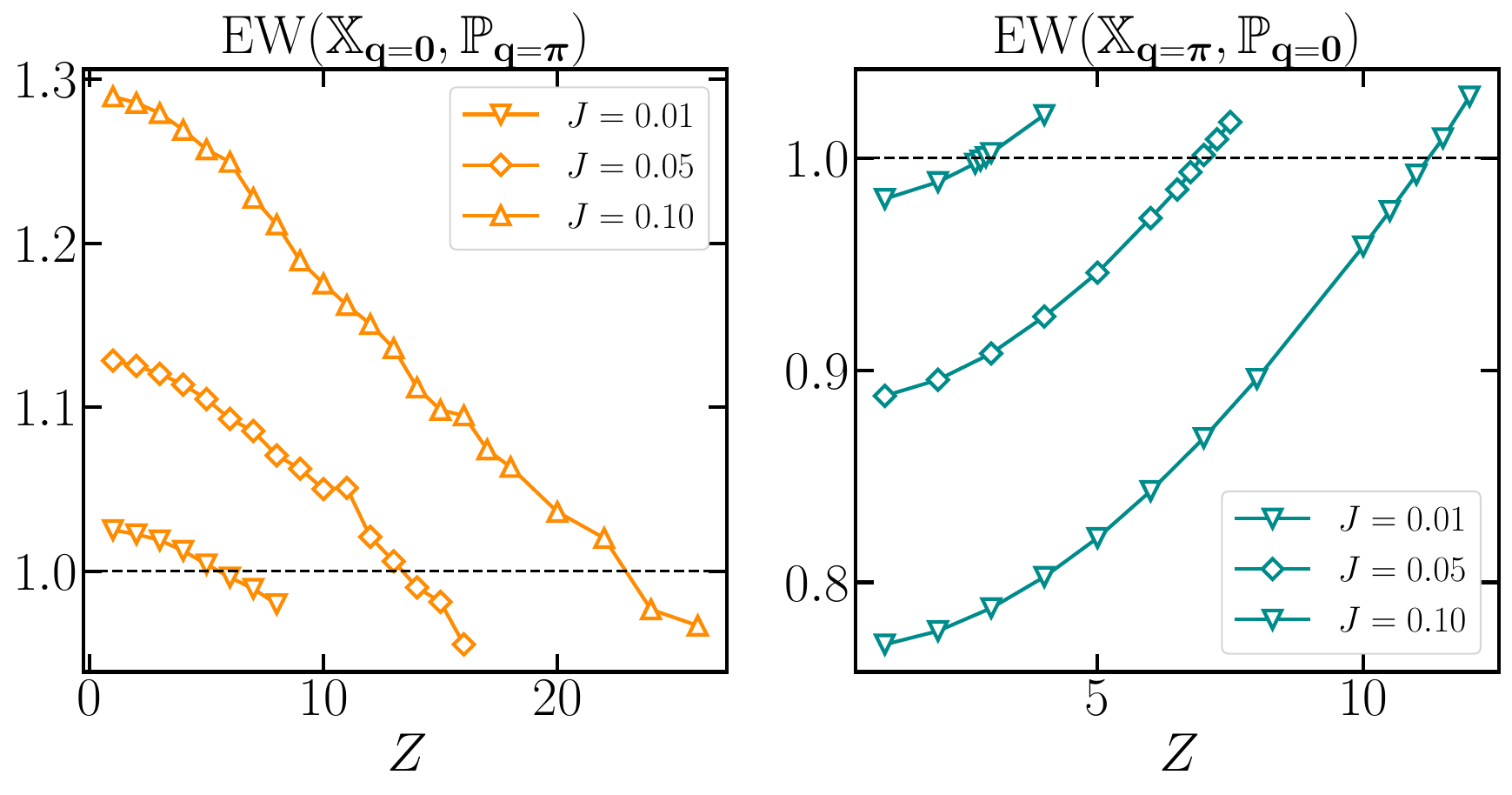}
\caption{Witnesses ${\rm EW}(\mathbb{X}_{\bq=\bs{0}},\mathbb{P}_{\bq=\bs{\pi}}) $ (left)
and ${\rm EW}(\mathbb{X}_{\bq=\bs{\pi}},\mathbb{P}_{\bq=\bs{0}}) $ (right) as a function of $Z$ and different values of $Z$. The horizontal lines indicates the separability 
bound. The color code matches the colors used in the main text 
for the ferroelectricity-induced (darkcyan) and ferroelectric-induced 
(orange) entanglement detection regions.}
\label{fig:supp_fig2b}
\end{figure}

We combine the fluctuations to construct the witnesses 
for different values of $J$ and $Z$ and map the detection phase diagram
by identifying the regions in which the witnesses fulfill the entanglement 
criterion, Eq. (6) in the main text. In Fig.~\ref{fig:supp_fig2b} 
we show the witnesses for different values of $J$ and $Z$. 
The crossing point of the separability bound is used to build 
the entanglement detection phase diagram reported in the main text.

\subsubsection{Fluctuation in a toy model of two minimally coupled quantum oscillators}\label{app:toy}
\begin{figure}[h]
\includegraphics[width=0.4\columnwidth]{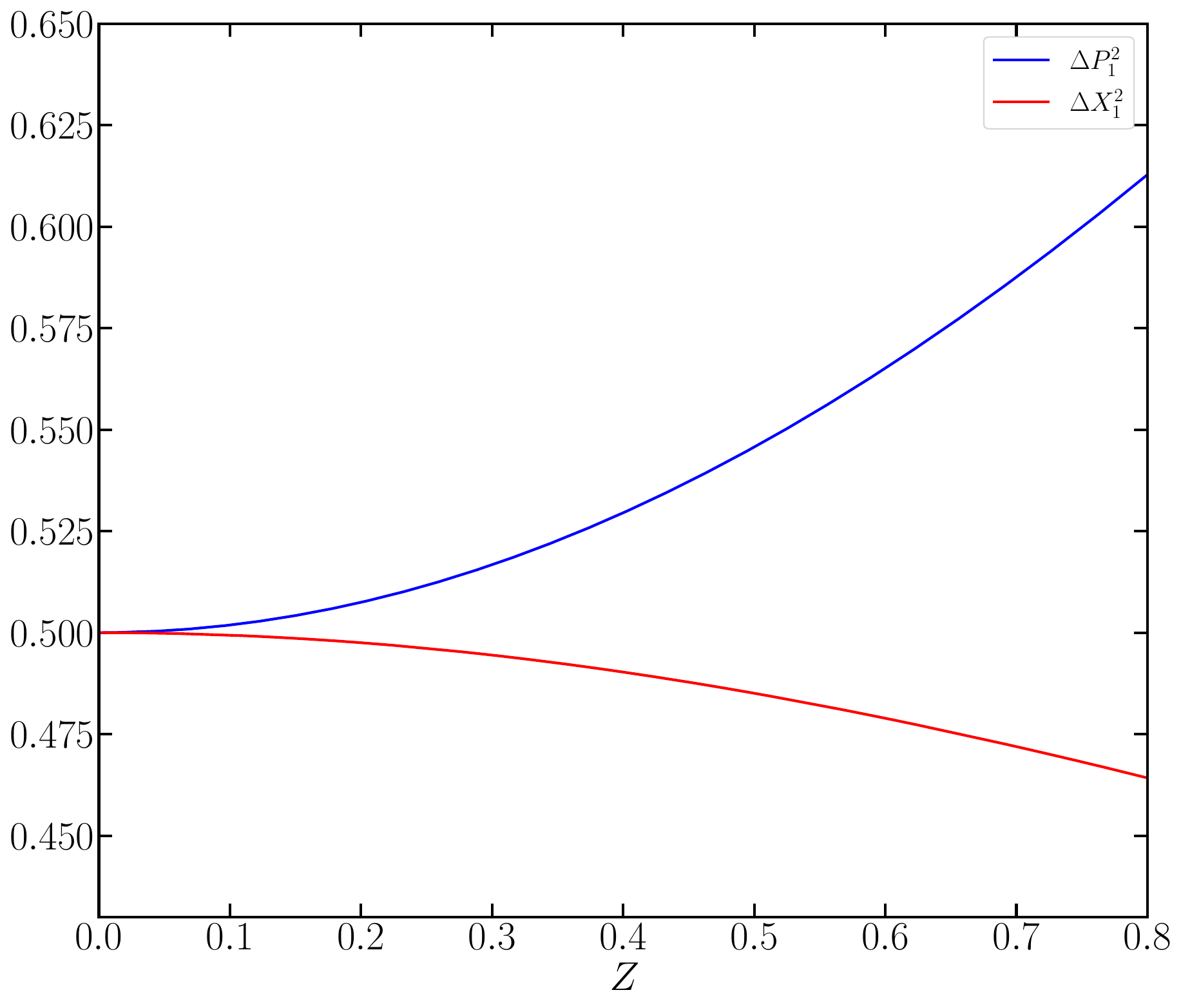}
\caption{Enhancement/suppression of the momentum/position fluctuations 
as a function of the coupling parameter, in the toy model in Eq.~\eqref{eq:Hamiltonian_toy}.}
\label{fig:supp_fig3}
\end{figure}

In order to understand the behavior of the position and momentum fluctuations
as a function of the light-matter coupling, we build a toy-model of two 
minimally coupled quantum oscillators.
We introduce two sets of conjugate variables, $(X_1,P_1)$  and $(X_2,P_2)$,
representing, respectively, the dipoles and the electromagnetic field. 
Here, $X_2$ plays the role of the vector potential operator, and $P_2$ the electric field operator.
We therefore write a minimally coupled Hamiltonian akin to the 
full Hamiltonian in Eq.~\eqref{eq:H_light_matter_full}
\begin{equation}
H = \frac{1}{2}(P_1+ Z X_2)^2 + \frac{1}{2} X_1^2 + \frac{P_2^2}{2} + \frac{X_2^2}{2}.
\label{eq:Hamiltonian_toy}
\end{equation}
We diagonalize the Hamiltonian in Eq.~\eqref{eq:Hamiltonian_toy} and compute the 
position and momentum fluctuations of the dipole operators.
In Fig.~\ref{fig:supp_fig3} we show that the toy model reproduces 
the enhancement/suppression of the momentum/position fluctuations
discussed in the main text.

\end{document}